\documentclass[11pt]{article}
\usepackage{amssymb, graphicx}
\newcommand\arcsec{\mbox{$^{\prime\prime}$}}

\begin{document}

\title{Cosmic Ray diffusion near the Bohm limit in the Cassiopeia~A
supernova remnant}

\author{M.\ D.\ Stage\\ 
\it{Five College Astronomy Department, University of Massachusetts}\\
\it{710 North Pleasant Street, Amherst, MA, 01003-9305}\\
and\\
G.\ E.\ Allen, J.\ C.\ Houck, and J.\ E.\ Davis\\ 
\it{MIT, Kavli Institute for Astrophysics and Space Research,}\\
\it{77 Massachusetts Avenue, Cambridge, MA, 02139-4307}
}

\maketitle

\bf Supernova remnants(SNRs) are believed to be the primary location of the
acceleration of Galactic cosmic rays, via diffusive shock (Fermi)
acceleration. Despite considerable theoretical work the precise details are
still unknown, in part because of the difficulty in directly observing
nucleons that are accelerated to TeV energies in, and affect the structure
of, the SNR shocks.  However, for the last ten years, X-ray observatories
{\it ASCA}, and more recently {\it Chandra}, {\it XMM-Newton}, and
{\it Suzaku} have made it possible to image the synchrotron emission at keV
energies produced by cosmic-ray {\it electrons} accelerated in the SNR
shocks. In this article, we describe a spatially-resolved spectroscopic
analysis of {\it Chandra} observations of the Galactic SNR Cassiopeia A to
map the cutoff frequencies of electrons accelerated in the forward shock.
We set upper limits on the diffusion coefficient and find locations where
particles appear to be accelerated nearly as fast as theoretically possible
(the Bohm limit).

\rm Supernova remnants (SNRs) have been established as the leading candidate
for the acceleration of cosmic rays.\cite{Baade34}\cite{GiSy64} It has been
shown that the mechanism of diffusive shock acceleration in SNR shocks
coupled with some understanding of Galactic transport effects can in theory
produce the observed power-law spectrum of cosmic
rays.\cite{Krymskii1977}-\cite{Ellison2000}.  This model works at least up
to the ``knee'' of the cosmic ray spectrum near $5 \times 10^{15}\;$eV, and
possibly all the way to the ``ankle'' near $3 \times
10^{18}\;$eV.\cite{Biermann2001} The charged particles scatter off of
irregularities in the magnetic field, increasing their momentum by a
fraction of the shock velocity $v/c$ with each round-trip shock
crossing.\cite{Biermann2001} Theoretical work in the last several years has
suggested that the process is significantly
nonlinear.\cite{Ellison2000}-\cite{Bell2005}
Higher energy particles have larger diffusion lengths and therefore sample a
greater change in velocity and compression ratio across the shock than 
lower energy particles.  This effect introduces a decrease in the particle
distribution's spectral index and a flattening of the spectrum with
increasing energy.\cite{Ellison2000} A smoothing effect on the structure of
the shock results, predominantly caused by the ions; electrons are almost
``test particles.'' At high energies the electron spectrum is expected to
have the same (curved) spectral shape as the proton
spectrum.\cite{Ellison2000} 

This correspondence of electron and proton spectra is significant because it
is extremely difficult to directly observe the acceleration of cosmic-ray
protons in SNR shocks. Their presence is inferred indirectly by detecting
$\pi^0$ decay $\gamma$-rays produced from collisions of the cosmic rays with
gas particles in the ambient medium. Atmospheric Cerenkov telescopes have
observed gamma rays from some supernovae: Cas A with
HEGRA\cite{Pul99}\cite{Aharonian2001}, RX~J1713.7-3946 (G347.3-0.5) with
CANGAROO\cite{Muraishi2000}\cite{Enomoto2002} and HESS\cite{Aharonian2004},
RX~J0852.0-4622 with HESS\cite{Aharonian2005b} and
CANGAROO-II\cite{Katagiri2005}, and G~0.9+0.1 with
HESS\cite{Aharonian2005a}. Current technology limits the spatial resolution
of these instruments to arcminutes and remnants the size of Cas A may appear
point-like.  In addition, TeV gamma rays can also be produced by
inverse-Compton scattering of cosmic microwave background photons with the
accelerated cosmic-ray electrons in SNRs.  Consequently, it is difficult to
precisely and unambiguously image proton and ion acceleration in the SNR. In
contrast, the electrons which are accelerated to TeV energies produce
synchrotron radiation at keV energies which is relatively easily imaged at
arcsecond resolution.  The electrons are not expected to be accelerated all
the way to ``knee'' energies, as at some point the synchrotron losses equal
the accelerative gains on shock crossing and the spectrum cuts off. However,
it is still possible to set a lower limit on the efficiency of the
acceleration.

An analysis of {\it ASCA} observations of SN~1006 provided the first 
evidence of X-ray synchrotron emission in an SNR.\cite{Koyama1995}.  The
approximately arcsecond spatial resolution over much of the
energy band (0.5 - 10 keV) of the {\it Chandra} High Resolution Mirror
Assembly--the best of any X-ray telescope which has flown\cite{Weiss2002}
--has enabled detailed studies.  For example, Lazendic et
al.\cite{Lazendic2004} studied {\it Chandra} observations of G347.3-0.5 to
localize cosmic ray acceleration in filaments.  Since the X-ray spectra of
the youngest Galactic supernova remnants (Cas A, Tycho, and Kepler) are
dominated by thermal emission, the ability to spatially resolve the
nonthermal contribution is especially important.  Using {\it Chandra}
observations, Warren et al.\cite{Warren2005} demonstrated that acceleration
in Tycho occurs at the forward shock.  

{\it Chandra's} Advanced CCD Imaging Spectrometer's (ACIS) spectral
resolution ($E/ \Delta E \approx 10-60$) allows extraction of the spectrum
of the emission. The combined capabilities of the mirror and CCDs allows us
to determine if a nonthermal spectrum is present and where it is located.
We:

1) developed a method to precisely identify the regions primarily radiating
nonthermal X-ray synchrotron emission in a remnant dominated by thermal
X-ray radiation from supernova ejecta; 

2) fit the spectra of thousands of individual regions dominated by
nonthermal emission in Cas A with a realistic synchrotron model, to map the
synchrotron cut-off frequency around the outer shock;

3) determined the ratio of the estimated upper limit on the diffusion
coefficient, $\bar\kappa$, in each region to the Bohm coefficient, $\kappa_{\rm
B}$, and thereby mapped the lower limit of the efficiency of acceleration in
a SNR. In several areas, diffusion appears to be occurring near the Bohm
limit. The maximum efficiency occurs in the northeast outer shock, just to
the side of the jet, where $\bar\kappa/\kappa_{\rm B} \le $ 2.1.

\section*{\bf \small Mapping the spectral properties of the Cas A SNR}
Exploiting the full capabilities of {\it Chandra's} spatial resolution,
ACIS's spectral resolution, the scripting functionality of the Interactive
Spectral Interpretation System (ISIS)\cite{Houck2000}, and distributed
computation, we produced high quality, high resolution maps of the spectral
parameters of the remnant.  We used direct spectral fitting, not energy
selection methods.  This process allowed us to map parameters such as
temperature and interstellar absorption column, which is not possible with
other methods. 

\begin{figure} 
\includegraphics[width=4.9in]{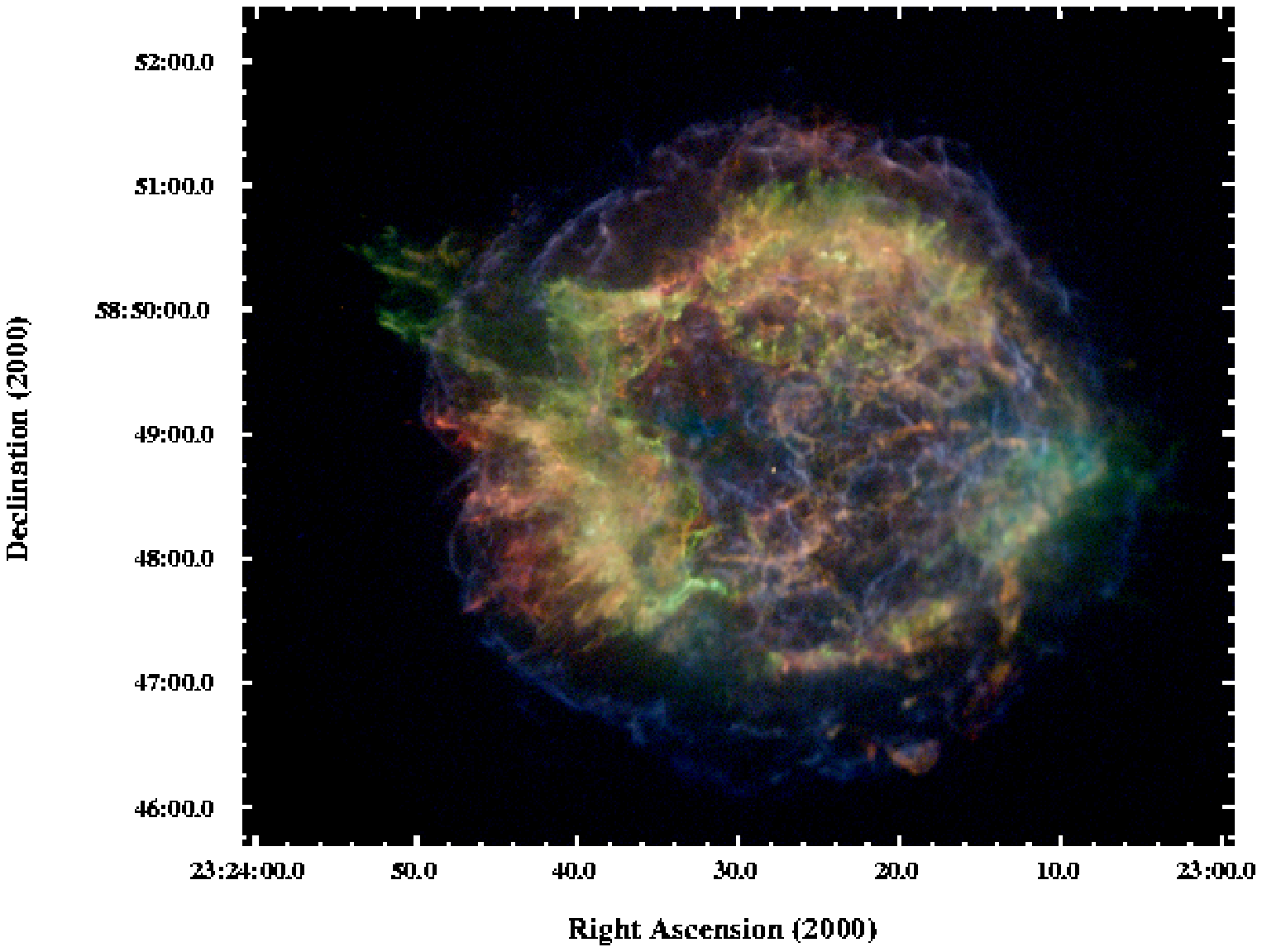} 
\caption{The Cassiopeia A supernova remnant as observed by {\it Chandra}.
Red: 0.5 - 1.5 keV: O-K, Ne-K, Mg-K, and Fe-L emission. Green: 1.5-2.5 keV:
Si-K and S-K.  Blue: 4.0-6.0 keV: high energy continuum (some energy ranges
have been excluded, including Fe-K).  This energy-colored photon counts
image was created from 1.08 Ms of ACIS data. The colors are
individually log-scaled to bring out the fine structure in the remnant.  The
bulk of the X-ray photons are thermal radiation from the ejecta which has 
passed through the hot reverse shock.  Note in particular the wispy blue
filamentary structure of the forward shock, dominated by high-energy
synchrotron emission from electrons accelerated at the forward shock. }
\label{fig:casa}
\end{figure}

We used the nine individual pointings from the million second observation of
Cas A\cite{Hwang2004} and two additional archival observations to take the
deepest look to date at a SNR with {\it Chandra}.  Figure \ref{fig:casa}
shows a combined counts image color-coded by energy.  To perform the
spatially-resolved spectral analysis, we divided the sky around Cas A into a
square grid of sky-coordinate region centers separated by approximately
1\arcsec~(2 ACIS pixels).  Cas A was covered by 104,393 regions. Each region
was adaptively sized to contain a minimum of 10,000 counts. The region
sizes ranged from 1\arcsec~$\times$~1\arcsec{} to 7\arcsec~$\times$~7\arcsec{}
on source. Regions were allowed to overlap; therefore, fits to
regions larger than 1\arcsec~$\times$~1\arcsec{} are not independent.

The position and roll angle of Cas A on the CCD varies in the eleven
pointings.  We applied standard processing and filtering to the data for
each pointing and then some custom techniques.\cite{Davis2005} We
re-projected the data to a common sky-coordinate plane, but maintained
separate files for each pointing. To speed the later extraction of spectra,
we histogrammed the spectral data and determined the mean chip coordinates
and the exposure time for each sky location.  This allowed us to separate
the slowly varying effect of the effective area, which we calculated on a 32
pixel by 32 pixel grid, and the quickly varying effect of bad pixels and the
exposure time, which we calculated on a pixel to pixel basis.  Since our
modified effective area depends only on chip location, we used one library
of the effective area function for all of the pointings corresponding to a
given observation epoch. Similarly, we pre-computed a library of energy
response matrices on the same 32 pixel by 32 pixel grid.  

Using ISIS, the same spectral model was fit jointly to the set of eleven
spectra for each region.  Each spectrum was associated with its own
effective exposure, response functions, and optionally, background spectrum
extracted from the same pointing.  The background spectrum was not
subtracted from the source spectrum, but was instead added to the model. The
background is dominated by photons from Cas A in the wings of the point
spread function of the telescope and in the CCD readout streak.  For
computational simplicity, the background for every source spectrum was drawn
from the same representative sky region. 

\begin{figure}
\includegraphics[width=4.9in]{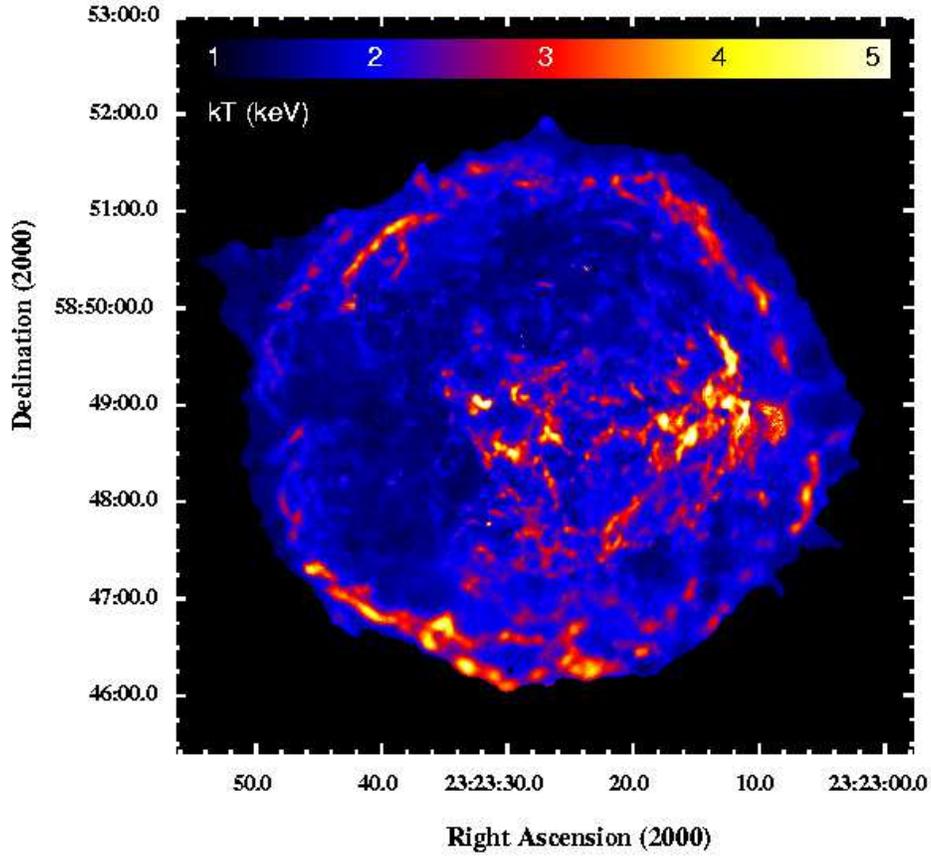} 
\caption{Cas A bremsstrahlung temperature map.  The image is of the fit
bremsstrahlung continuum from the Phase 1 analysis of the remnant.  We fit
the spectrum of 104,393 regions with a model that includes 15 Gaussian
lines, interstellar absorption, and a thermal bremsstrahlung continuum. Note
the correlation between the high-temperature features in this image and the
blue, high-energy continuum-dominated filaments in Figure \ref{fig:casa}.
The anomalously high temperatures in this image indicate nonthermal emission
in the SNR, which is likely synchrotron radiation from electrons accelerated
to TeV energies.} \label{fig:ktmap}
\end{figure}

\begin{figure}
\includegraphics[angle=270,width=4.9in]{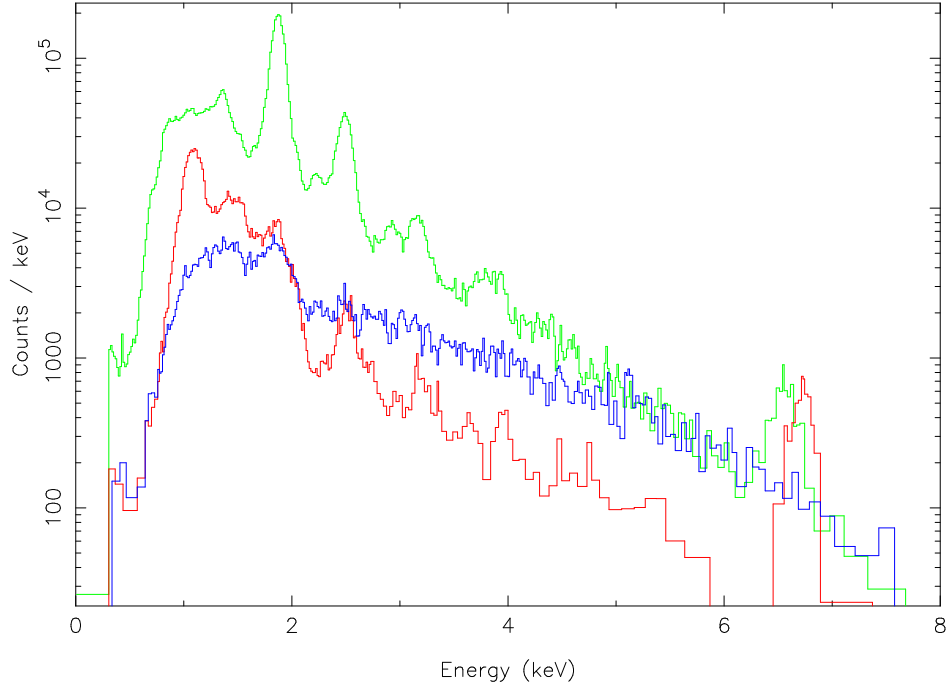} \caption{Comparison
of a continuum-dominated spectrum (blue) from a 2\arcsec~by 2\arcsec~region
on the northeast shock ($\alpha$=23:23:42.2, $\delta$=58:50:26.4) shows it
has a different spectral shape than spectra from same-sized regions
dominated by line emission (Si, S, Fe, O, Ne, Mg, Ar, Ca). The green
spectrum is taken from a region in the northern lobe ($\alpha$=23:23:52.1,
$\delta$=58:50:33.3) featuring strong K-line emission from silicon and
sulfur.  The red spectrum is from the fingerlike projections in the
southeast lobe ($\alpha$=23:23:44.0, $\delta$=58:48:04.7) dominated by iron
L and K emission.  The exposure time is 1.08 Ms, and for display the spectra
are binned to contain at least 10 counts per bin.} \label{fig:regions}
\end{figure}

The analyses were performed in two phases. In Phase 1, we used a simple
fitting function of fifteen Gaussian lines for specific O, Ne, Mg, Si, S, 
Ar, Ca, Ti and Fe line complexes (representing He-like K$_\alpha$ for all,
He-like K$_\beta$ for Si and S, H-like K$_\alpha$ for Si, and L lines for
multiple Fe ions); an interstellar absorption column; and a bremsstrahlung
continuum. This particular model was fit without background spectra (because
it fits the background) to the 104,393 regions covering the entire remnant.
The joint best-fit spectral parameters were recorded for each region and
used to create FITS images of the spectral properties. Figure
\ref{fig:ktmap} shows the fit temperatures of the bremsstrahlung continuum.
Structures concentrated at the outer edges and in part of the center show a
significantly higher temperature than the rest of the remnant.  As shown in
Figures \ref{fig:ktmap} and \ref{fig:regions}, these regions show little
evidence of line emission and are correlated with anomalously high
absorption column values and with the high energy (4-6 keV)
continuum-dominated filaments in Figure \ref{fig:casa}. There appears to be
some genuine elevation of the absorption column in the western region of the
remnant, where there is evidence that the remnant is obscured by a molecular
cloud\cite{KRA96}. At the outermost edges, the filaments have been
associated with the outer shock.\cite{Gotthelf2001}. Collectively, these
clues suggest that the model used for the first phase of the analysis is
inappropriate in the anomalously high temperature regions because the
emission is nonthermal.  While some of the high energy X-ray emission (10-60
keV)\cite{Allen1997} from Cas A may be produced by nonthermal bremsstrahlung
\cite{Laming2001}-\cite{Vink2003}, there is general
agreement that the X-ray emission observed from forward-shock filaments is
produced by synchrotron emission.  Therefore, in Phase 2, we fit the spectra
of regions identified by this method with a model composed of realistic
synchrotron emission, an absorption column, and a background spectrum.

\section*{\small \bf Creating a cut-off frequency map}

Having found a means to identify the regions dominated by nonthermal
emission in a remnant whose X-ray emission is globally dominated by thermal
emission, we studied the properties of particle acceleration.  We used the
kT results to pick 10,857 regions--the top 10\% of fitted temperatures 
(kT~$>$~2.6~keV).  This criterion clearly selected areas centered on and
fully containing the outer filaments but not excessive surrounding areas or
isolated peaks.  Since the fit cut-off frequencies span a range of two
orders of magnitude, we believe we found all of the regions of interest
without introducing selection effects in the distribution of detected
cut-offs.

Our group developed a synchrotron spectrum model in which the effects of
spectral ``curvature'' from the back-pressure of the accelerated cosmic rays
are modelled using an electron spectrum whose spectral index is a linear
function of the logarithm of the momentum.\cite{Houck2006}.  The core of the
model is the electron distribution function,

\begin{equation}
 \frac{dn}{dpc} = A_e(\frac{pc}{{\rm GeV}})^{-\Gamma+a\;{\rm
log_{10}}(\frac{pc}{_{\rm GeV}})}\;e^{({\rm GeV}-pc)/\epsilon}
\end{equation}

\noindent in which $n$ is the electron number density, $c$ is the speed of
light, $p=\gamma m v$ with $m$ the electron mass and $v$ the velocity of a
particle, $A_e$ is the number density at $p=$1 GeV$/c$, $\Gamma$ is the
differential spectral index at $p=$1 GeV/$c$, $\epsilon$ is the exponential
cut-off (maximum) energy, and $a$ is the spectral ``curvature.'' The
parameters of the synchrotron model are the normalization $N_s$ (a function
of $A_e$, distance, and the emitting volume), the total magnetic field
$B_{\rm tot}$, $\Gamma$, $a$, and $\epsilon$.  

The cut-off energy $\epsilon$, the normalization of the spectrum and the
value of an interstellar absorption column were allowed to be free
parameters.  We froze the curvature parameter to a value of 0.06 based on
fits to radio and infrared synchrotron data for selected regions of
Cas~A\cite{Jones2003}.  A value of $a= 0.05$ was obtained by our group in
fits to SN1006\cite{AllenHouckSturn}. As there are no published values for
the radio spectral index of Cas A at the outer shocks, which are at the edge
of a diffuse plateau of radio emission, we set $\Gamma = 2.54$, using the
average radio spectral index for the entire remnant.\cite{Baars1977}
Comparison with the index in the plateau region just inside the outer X-ray
filaments from unpublished radio spectral index maps suggests this value is
reasonable.  Small variations in the value of $\Gamma$ did not significantly
affect our final results.  Changing the value of $\Gamma$ by $\pm 0.1$
changed the cut-off frequencies to values within our confidence limits and
preserved overall trends.  

Some assume the widths of the filaments are limited by synchrotron losses
and estimate the strength of the magnetic field in the filaments of Cas~A to
be $\approx$ 100 $\mu$G\cite{Vink2003} to $\approx$ 500
$\mu$G\cite{Berez2004}. Others argue the sizes of the filaments in SNRs are
determined by the widths of regions of enhanced magnetic fields at the
forward shock, limited by strong turbulent damping of the field to physical
sizes of order $10^{16}-10^{17}$ cm\cite{Pohl2005}.  This range corresponds
to 0.2 to 2\arcsec{} at the distance of Cas~A, similar in size to the
observed width. The maximum frequency of emission from electrons is related
to the maximum energy and the magnetic field as

\begin{equation}
 \nu_{\rm c} = 1.26\times 10^{16}(\frac{\epsilon}{10\;{\rm
TeV}})^2(\frac{B_{\rm tot}}{{\rm 10\;\mu G}}){\rm Hz}.
\end{equation}

It may not be possible to uniquely determine $\epsilon$ and $B_{\rm tot}$
based on solely on analyses of the X-ray spectra.  Therefore, we froze
$B_{\rm tot} = 1000 \mu$G (the ``equipartition value'' for Cas
A\cite{Allen1997}).  As a result, the cut-off energy itself is not well
determined, but the cut-off frequency is well determined by the X-ray data. 

\begin{figure}
\includegraphics[width=4.9in]{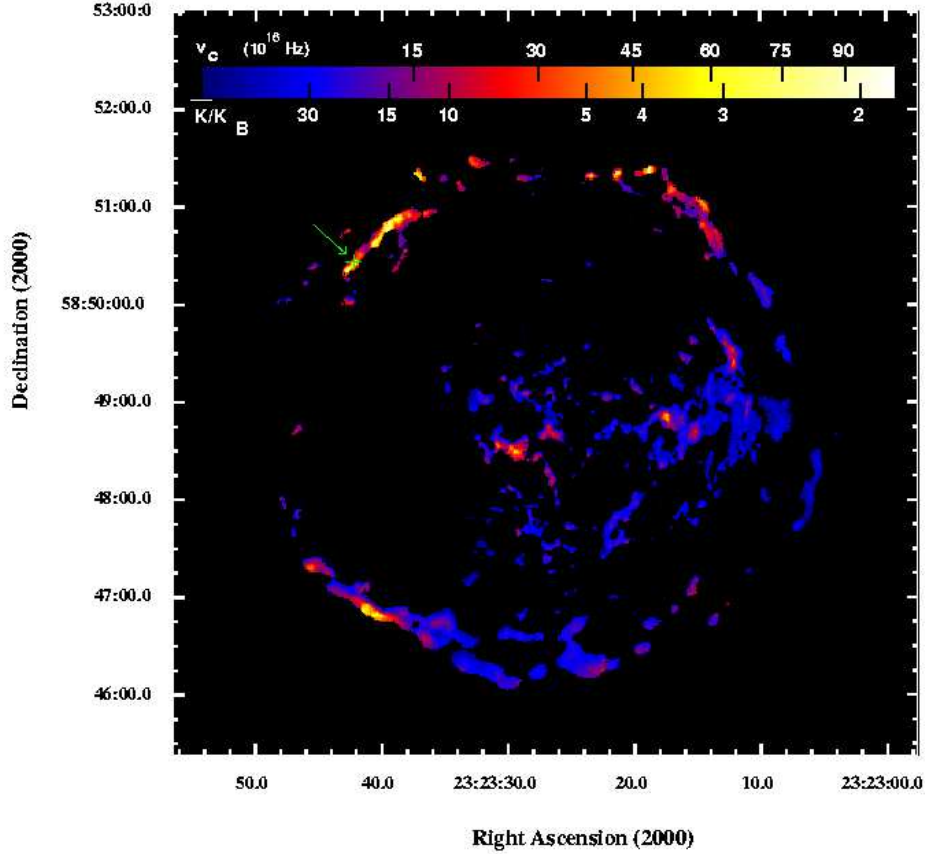} \caption{Synchrotron cut-off
frequencies and upper limits on the electron diffusion coefficient in the
Cassiopeia A supernova remnant.  A synchrotron model which includes
curvature in the electron distribution spectrum was used to fit the spectra
of 10,857 regions identified from the kT map as likely sites of nonthermal
emission (see Figure \ref{fig:spectrum}) because they have anomalously high
bremsstrahlung temperatures (i.e.\ kT~$>$~2.6~keV, Figure \ref{fig:ktmap}).
The fitted cut-off frequencies and estimates of the shock velocity and
compression ratio are used to determine an upper limit on the ratio of the
electron diffusion coefficient ($\bar\kappa$) to the Bohm diffusion
coefficient ($\kappa_{\rm B}$).  The highest fit cutoff value occurs at the
2\arcsec~$\times$~2\arcsec{} region indicated by the green cross and arrow;
the fit is shown in Figure \ref{fig:spectrum}.  A ratio close to 1, as seen
in the yellow and white regions of the map, indicates that acceleration must
be occurring almost as fast as theoretically possible (in the Bohm limit). }
\label{fig:cut-offmap}
\end{figure}

\begin{figure}
\includegraphics[angle=270,width=4.9in]{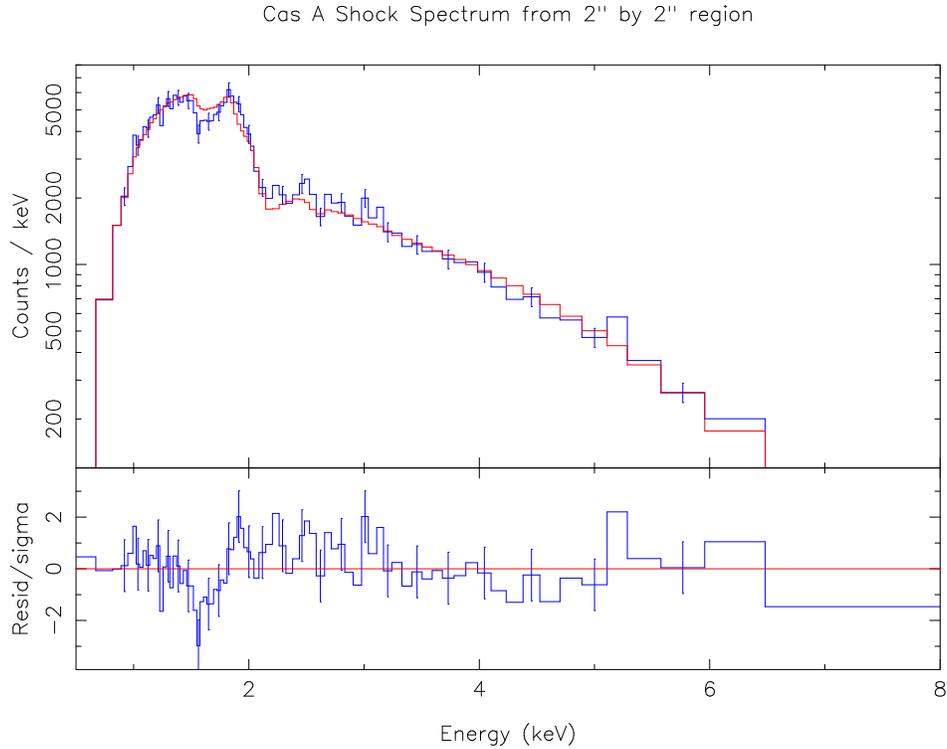} \caption{The
nonthermal continuum spectrum of the 2\arcsec~by 2\arcsec~region on the
northeast shock ($\alpha$=23:23:42.2, $\delta$ = 58:50:26.4) which shows the
highest cut-off frequency, 9 $\times 10^{17}\;$Hz.  The blue histogram is
the combined data; the red histogram is the model plus combined background.
The apparent (thermal) structure between 1 and 3 keV is primarily
background.  Only every third error bar has been plotted to improve clarity.
The exposure time is 1.08 Msec and for fitting the spectrum has been binned
to contain a minimum of 100 counts per bin.} \label{fig:spectrum}
\end{figure}

We present our map of $\nu_{\rm c}$ in Figure \ref{fig:cut-offmap}.  Cut-off
frequencies in the forward shock vary at least from 5 $\times 10^{16}\;$Hz
to 9$\times10^{17}\;$Hz.  The 90\% confidence intervals have been computed
for all of the cut-off frequencies.  The average uncertainties range from
about $\pm$ 20\% at 5 $\times 10^{16}\;$Hz to $\pm$ 40\% at
9$\times10^{17}\;$Hz. Therefore, the cut-off energy and/or the magnetic
field vary considerably from one region to another.  Variations in the
cut-off frequency have been reported for a few other 
remnants\cite{Lazendic2004},\cite{Pannuti2003}-\cite{Rothenflug2004}. The
spectrum of a 2\arcsec~$\times$~2\arcsec{} region along the shock in the 
northeast is shown in Figure \ref{fig:spectrum}.  This region has the
highest cut-off frequency (9$\times 10^{17}$ Hz).

\section*{\small \bf Constraining the electron diffusion coefficient}

As recently reviewed by Vink\cite{Vink2006}, some analyses are based on the
assumption that particle acceleration is efficient: the particle diffusion
coefficient is as small as the Bohm limit\cite{Volk2005}\cite{Bamba2005}.
Here, we use the spectral results to constrain the efficiency of
acceleration of the highest energy electrons at each location dominated by
synchrotron radiation.  

Assuming that electrons at the cut-off energy cannot experience synchrotron
 losses in excess of accelerative gains in one cycle through the shock, one
 can derive an upper limit on the ratio of the electron diffusion
 coefficient to the Bohm coefficient.\cite{AllenHouckSturn}.  The ratio is a
 function of the cut-off frequency, the shock velocity $u_{1}$ and the
 compression ratio $r$. At $E=\epsilon$, $\kappa_{\rm B}=\epsilon/{3eB}$ in
 MKSA and  

\begin{eqnarray}
\frac{\bar{\kappa}}{\kappa_{\rm B}} & \le & 
  \frac{9 \epsilon_{0} m c}{2 e^{2}}
  \frac{f u_{1}^{2}}{\nu_{\rm c} \sin \theta}
  \label{eqn_diff_coeff_ratio1} \\
& = & 2.08
  \left( \frac{f}{0.15} \right)
  \left( \frac{u_{1}}{5000~{\rm km}~{\rm s}^{-1}} \right)^{2}
  \left( \frac{\nu_{\rm c}}{9 \times 10^{17}~{\rm Hz}} \right)^{-1}.
  \label{eqn_diff_coeff_ratio2}
\end{eqnarray}

\noindent Here $\epsilon_{0}$ is the permittivity of free space, and $f=
(r-1)/((r)(r+1))$.

To constrain the electron acceleration efficiency, we used a value of
5000~km/s for the shock velocity of Cas A\cite{DeLaney2003}, and assumed the
case of a strong, unmodified shock ($f=0.15, r=4$).  For the highest cut-off
frequencies on the outer shock of Cas A, we obtained $\bar\kappa/\kappa_{\rm
B}$ = 2.1$\stackrel{+0.9}{_{-0.7}}$.  The variations on the outer
rim range from 0.5 to 9$\times 10^{17}$ Hz, corresponding to
$\bar\kappa/\kappa_{\rm B}\sim$ 36 to 2.  Our upper limits are smallest in
certain parts of the north, northeast, and southeast shocks, strongly 
suggesting efficient particle acceleration occurs in at least these regions.

\section*{\bf \small Conclusions}

We developed a method to analyze the emission from extended X-ray sources.
We combined multiple data sets retaining pointing-specific calibrations and
backgrounds, automated spectral extraction, and fully exploited the spatial
and spectral capabilities of {\it Chandra}.  Analyzing the emission from the
young Cassiopeia A supernova remnant, we discovered the apparent
bremsstrahlung temperature ($kT$) of regions can be used to identify
individual regions which are dominated by nonthermal emission in a SNR
globally dominated by thermal emission. For Cas A, we found that the
synchrotron emission is clearly correlated with the forward shock, as well
as some areas in the center, which may or may not be the forward shock seen
in projection. Fitting the nonthermally dominated regions with a synchrotron
model, we mapped the cut-off frequency of the synchrotron spectrum
associated with the highest energy electrons accelerated in the forward
shock.  From the cut-off frequency and shock velocity, we constrained the
efficiency of the acceleration of cosmic-ray electrons. 

We present the first map of the upper limits on the particle diffusion
coefficients in a SNR. Our results suggest that at least in several
locations, acceleration occurs nearly as fast as theoretically possible (the
Bohm limit). In fact, electrons could also be diffusing in the Bohm limit in
other parts of the shock if the shock velocity is lower; if the compression
ratio is greater than 4---an enhancement quite possible in a modified shock;
or if a process other than synchrotron radiation limits the maximum energy.
TeV $\gamma$-rays have been detected from
Cas~A\cite{Pul99}\cite{Aharonian2001}. Modeling suggests these could be
formed either by the prompt decay of $\pi^0$'s created in collisions of
cosmic-ray protons accelerated in the SNR with ambient gas; or, by
inverse-Compton scattering of cosmic microwave background photons off of the
accelerated electrons. Although TeV telescopes cannot yet localize the
$\gamma$-ray source position with arcsecond accuracy, or determine
unambiguously the fraction of $\pi^0$ and inverse-Compton produced
$\gamma$-rays from the spectrum, the expectation that cosmic-ray electrons
and protons are accelerated together means that electron synchrotron
emission is an important tracer of the particle acceleration.  Detailed maps
of the X-ray synchrotron emission, such as those we created with {\it
Chandra}, are rapidly improving our understanding of the big picture of
cosmic-ray acceleration in SNRs.

\section*{Acknowledgements} 

MDS and GEA thank Tracey Delaney and Lawrence Rudnick for sharing
unpublished radio spectral index data of Cas A to help us in
determining a reasonable choice of $\Gamma$ for the forward shock.
This work was supported in part by NASA LTSA grant NAG5-9237 and the
Five College Astronomy Department Fellowship program.

\clearpage

\clearpage

\clearpage

\clearpage

\clearpage

\end{document}